\documentclass[sigconf]{acmart}

\copyrightyear{2022}
\acmYear{2022}
\setcopyright{acmcopyright}
\acmConference[CIKM '22] {Proceedings of the 31st ACM International Conference on Information and Knowledge Management}{October 17--21, 2022}{Atlanta, GA, USA.}
\acmBooktitle{Proceedings of the 31st ACM International Conference on Information and Knowledge Management (CIKM '22), October 17--21, 2022, Atlanta, GA, USA}
\acmPrice{15.00}
\acmISBN{978-1-4503-9236-5/22/10}
\acmDOI{10.1145/3511808.3557668}

\AtBeginDocument{%
  \providecommand\BibTeX{{%
    \normalfont B\kern-0.5em{\scshape i\kern-0.25em b}\kern-0.8em\TeX}}}

\begin{document}
\title{ Personalized Federated Recommendation via Joint Representation Learning, User Clustering, and Model Adaptation}
\author{Sichun Luo}
\affiliation{%
\department{Department of Computer Science}
  \institution{City University of Hong Kong}
  \city{Hong Kong}
  \country{China}
}
\affiliation{%
  \institution{City University of Hong Kong Shenzhen Research Institute}
  \city{Shenzhen}
  \country{China}
}
\email{sichun.luo@my.cityu.edu.hk}

\author{Yuanzhang Xiao}
\affiliation{%
\department{Hawaii Advanced Wireless Technologies Institute}
  \institution{University of Hawaii at Manoa}
  \city{Honolulu}
  \state{HI}
  \country{USA}
}
\email{yxiao8@hawaii.edu}

\author{Linqi Song}
\affiliation{%
\department{Department of Computer Science}
  \institution{City University of Hong Kong}
  \city{Hong Kong}
  \country{China}
}
\affiliation{%
  \institution{City University of Hong Kong Shenzhen Research Institute}
  \city{Shenzhen}
  \country{China}
}
\email{linqi.song@cityu.edu.hk}
\authornote{Corresponding author}

\begin{abstract}





Federated recommendation applies federated learning techniques in recommendation systems to help protect user privacy by exchanging models instead of raw user data between user devices and the central server. Due to the heterogeneity in user's attributes and local data, attaining personalized models is critical to help improve the federated recommendation performance. In this paper, we propose a Graph Neural Network based Personalized Federated Recommendation (PerFedRec) framework via joint representation learning, user clustering, and model adaptation. Specifically, we construct a collaborative graph and incorporate attribute information to jointly learn the representation through a federated GNN. Based on these learned representations, we cluster users into different user groups and learn personalized models for each cluster. Then each user learns a personalized model by combining the global federated model, the cluster-level federated model, and the user's fine-tuned local model. To alleviate the heavy communication burden,  we intelligently select a few representative users (instead of randomly picked users) from each cluster to participate in training. Experiments on real-world datasets show that our proposed method achieves superior performance over existing methods.
\end{abstract}



\maketitle

\section{Introduction}
Federated recommendation aims to help users filter out useful information while keeping users' personal data private. Following the principles in federated learning \cite{mcmahan2017communication}, federated recommendation exchanges recommendation models, instead of raw data, between user devices and the central server. This new distributed learning paradigm has found applications in content recommendations \cite{tan2020federated,ali2021federated}, mobile crowdsourcing task recommendations \cite{zhang2020pfcrowd,guo2020fedcrowd}, and autonomous driving strategy recommendations \cite{savazzi2021opportunities}.

Several classic recommendation algorithms have been extended to the federated setting, such as federated collaborative filtering \cite{ammad2019federated,minto2021stronger}, federated matrix factorization \cite{chai2020secure,li2021federated,du2021federated}, and federated graph neural networks (GNN) \cite{wu2021fedgnn}. However, these works have two drawbacks. First, they use the \textit{same} aggregated recommendation model for all the users, ignoring the heterogeneity of the users (e.g., non-IID data distribution, different levels of computing resources). Second, they require model exchanges between the server and \textit{all} the users for federated learning, incurring high communication costs.


To address the above issues, we propose a graph neural network based \textbf{Per}sonalized \textbf{Fed}erated \textbf{Rec}ommendation (PerFedRec) framework. PerFedRec learns user representations from user/item attributes and collaborative information (i.e., user-item interaction) via a GNN, and group similar users into clusters based on user representation. Then each cluster attains a cluster-level federated recommendation model, and the server attains a global model. Finally, each user combines its local model, the cluster-level model, and the global model to obtain a personalized model. Importantly, the representation learning, user clustering, and model adaptation are done jointly instead of independently.

In addition to personalization, our proposed PerFedRec framework alleviates the heavy communication burden by judiciously picking a few representative clients in each cluster in the training of the global model. This ``user dropout'' could save communication costs in critical scenarios, such as (autonomous) driving strategy recommendations where wireless bandwidth is limited and long time delay is prohibitive.   
To sum up, our contribution is threefold.

$\bullet$ We propose a joint representation learning, user clustering, and model adaptation framework in federated recommendation to achieve personalized recommendation, which adapts to the heterogeneity of users' local data and resources. We show that representations, learned by a GNN from users' collaborative and attribute information, are helpful to cluster similar users and learn personalized models. 
 
$\bullet$ We carefully pick a few representative users from each cluster to participate in training. This approach reduces communication costs, and is especially suitable for applications with limited bandwidth and low latency requirement.

$\bullet$ Our proposed approach improves the performance of state-of-the-art baselines on several real-world datasets.

\begin{figure*}[htbp!]
    \centering
    \includegraphics[width=0.55\textwidth,height=0.45\textwidth]{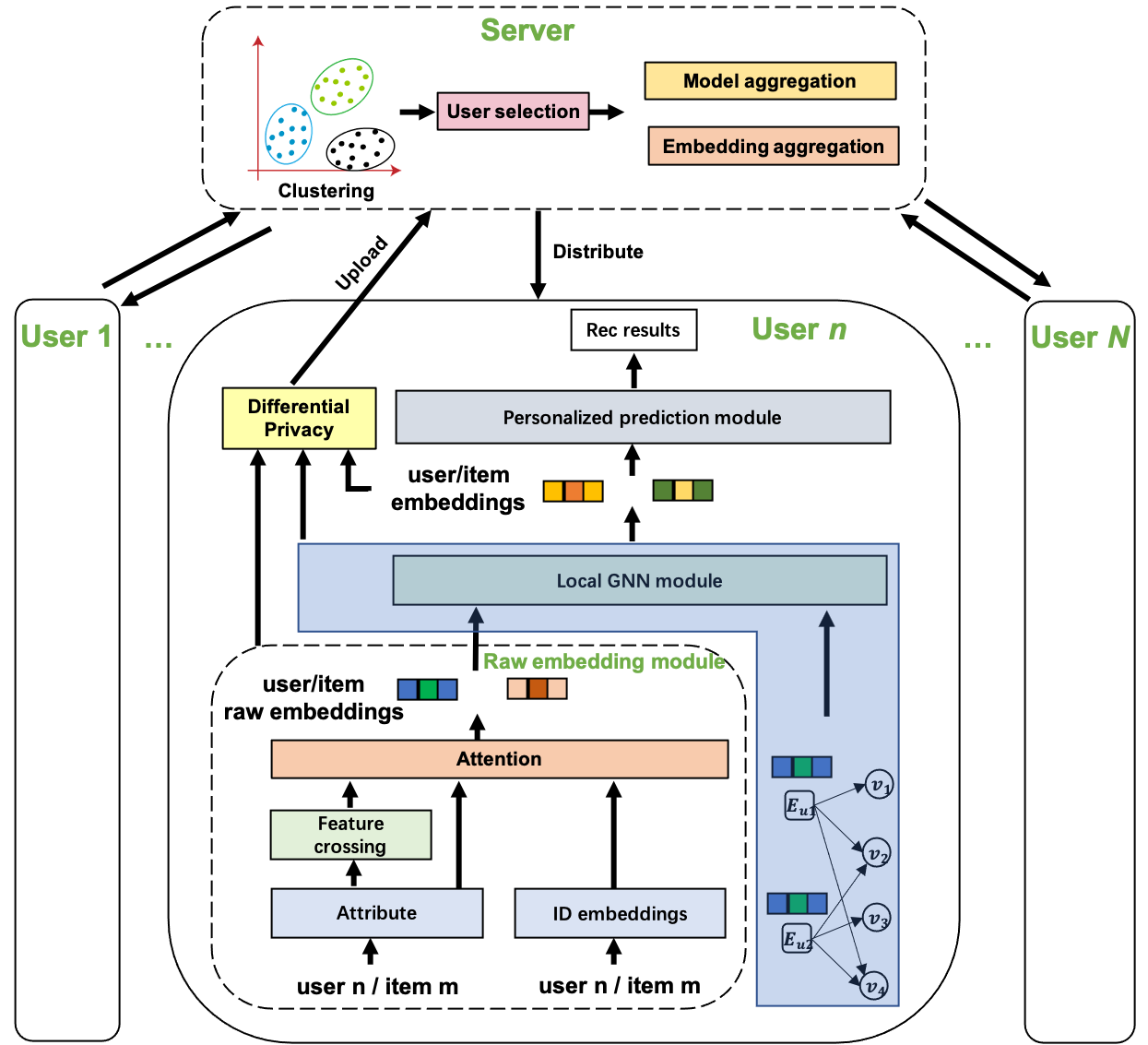}
    \caption{Illustration of the proposed personalized federated recommendation framework.}
    \label{overall}
\end{figure*}
\section{Problem Formulation}
We consider a federated recommendation system consisting of a central server and $N$ distributed users, each with a $d_{ua}$-dimensional attribute $u_n \in \mathbb{R}^{d_{ua}}$, $\forall n \in [N]$. There are $M$ items, each with a $d_{ia}$-dimensional attribute $v_m \in \mathbb{R}^{d_{ia}}$, $\forall m\in [M]$, to be recommended in the system. Each user has some historical interactions with items (e.g., rating the items). However, the users' historical interactions and attributes cannot be observed by the central server due to users' privacy concerns. Therefore, only recommendation models, instead of user data, can be exchanged between the server and the user devices. Under such constraints, the system aims to train personalized recommendation models for different users. 

\section{Our Proposed Framework}

Our proposed PerFedRec framework consists of a user-side end-to-end local recommendation network and a server-side clustering based aggregator. The overall architecture is shown in Fig.~\ref{overall}. 

\subsection{User-Side Local Recommendation Network}
Our proposed user-side local recommendation network has three modules: a raw embedding module, a local GNN module, and a personalized prediction module.

$\bullet$~{\bf Raw Embedding Module.} This module pre-processes user and item attributes. Via an attention mechanism, it combines attribute information with collaborative information (i.e., user-item interactions) to generate inputs to the local GNN module. Formally, the collaborative information for user $n$ and item $m$ is denoted by a $d$-dimensional ID embedding $E_{idu,n} \in \mathbb{R}^{d}$ and $E_{idi,m} \in \mathbb{R}^{d}$, respectively. These embeddings are initialized randomly and updated based on user-item interactions during training. The attributes of user $n$ and item $m$ are passed through a linear layer and a feature crossing layer to generate the attribute embeddings $E_{fcu,n} \in \mathbb{R}^{d}$ and $E_{fci,m} \in \mathbb{R}^{d}$, respectively:
\begin{equation}
    \begin{split}
      E_{fu,n} &= W_{fu,n}u_n+b_{fu,n}, \\
      E_{fi,m} &= W_{fi,m}v_m+b_{fi,m}, \\
      E_{fcu,n} &= \text{FeatureCrossing}(E_{fu,n}),\\
      E_{fci,m} &= \text{FeatureCrossing}(E_{fi,m}),
    \end{split}
\end{equation}
where $W_{fu,n},b_{fu,n},W_{fi,m},b_{fi,m}$ are network parameters of the linear layer, and $\text{FeatureCrossing}({\mathbf x}_0)$ is the feature crossing network that mixes the information in ${\mathbf x}_0$ across dimensions. The feature crossing network consists of $L$ feature crossing layers and the output $\mathbf{x}_{l+1} \in \mathbb{R}^{d}$ of layer $l+1$ is obtained by $\mathbf{x}_{l+1} = \mathbf{x}_0 \mathbf{x}_l^\top \mathbf{w}_l + \mathbf{b}_l + \mathbf{x}_l$, where $\mathbf{x}_{l} \in \mathbb{R}^{d}$ is the output of layer $l$, and $\mathbf{w}_l, \mathbf{b}_l \in \mathbb{R}^{d}$ are parameters of layer $l$. 

After getting these cross feature embeddings of attributes, we use an attention network to incorporate the attribute into the collaborative information:
\begin{equation}
    \begin{split}
      & E_{attu,n} = \text{Attention}(E_{fu,n},E_{fcu,n},E_{idu,n}),\\
      & E_{atti,m} = \text{Attention}(E_{fi,m},E_{fci,m},E_{idi,m}),
      \end{split}
\end{equation}
where $\text{Attention}()$ is the attention mechanism. One example of the attention mechanism is:
\begin{equation}
    \begin{split}
    \text{Attention}(\mathbf{x}_1,\ldots,\mathbf{x}_k) = 
    \sum_{i=1}^k \frac{\text{exp}(\text{tanh}(W_i \mathbf{x}_i + b_i))}{\sum_{i^\prime=1}^k \text{exp}(\text{tanh}(W_{i^\prime}\mathbf{x}_{i^\prime} + b_{i^\prime})) } \mathbf{x}_i.
    \end{split}
\end{equation}

We concatenate all the above embeddings as the raw embeddings for user $n$ and item $m$: $E_{rawu,n}$ and $E_{rawi,m}$.

Item embeddings are shared and updated iteratively among users via the server, while the user embeddings are kept locally due to privacy concerns. These raw initial embeddings are used to train the local GNNs. During the training process, the global item embeddings and local user embeddings will be updated.

$\bullet$~{\bf Local GNN Module.}
After getting all items' embeddings and the user's own embedding, each user needs the user-item interaction matrix to train the local GNN model. However, one difficulty is that user-item interaction information is kept private as local data and should not be shared among the server and other users.

In order to tackle such issues, we follow a similar idea as in \cite{wu2021fedgnn}, where each user uploads the privacy-protected embedding and the encrypted IDs (with the same encryption for all users) of the items that this user has interaction with to the server. Then the server sends encrypted item IDs and user embeddings back to all the users. To further reduce the communication cost, the server can just send back the encrypted item ID and the other users' embeddings to a user that has previously interacted with this item. Therefore, each user is able to get several users' embedding information together with the corresponding items, without revealing the identities of these users. In this way, each user could explore its neighborhood users and expand a local interaction graph.



The GNN module will output user $n$'s embedding $E_{u,n}$ and items' embeddings $\{E_{i,m}\}$: 
\begin{equation}
    \begin{split}
     & E_{u,n}, E_{i,m} = \text{LocalGNN}(\{E_{rawu,n}\},\{E_{rawi,m}\}). \\
     \end{split}
\end{equation}
These embeddings will be fed into the personalized prediction network for rating/preference predictions. 

Note that there are various choices for this plug and use GNN module, such as PinSage \cite{ying2018graph}, NGCF \cite{wang2019neural} and LightGCN \cite{he2020lightgcn}. 

$\bullet$~{\bf Personalized Prediction Module.}
To achieve personalized recommendation, our framework trains personalized recommendation models for each user. Let us denote by $\theta^{t}_{local,n}$ the raw embedding and GNN model trained by user $n$ at time step $t$. From the server side, we will also have a global federated model $\theta^{t}_{global}$ and a cluster-level federated model $\theta^{t}_{C(n)}$, where $C(n)$ is the cluster containing user $n$. We will describe how to obtain the global and the cluster-level models later. The personalized model combines these three models together via weights $\alpha_{n,1},\alpha_{n,2},\alpha_{n,3}$, which can be either hyperparameters or learnable parameters:
\begin{equation}
    \begin{split}
     & \theta^{t}_{n} = \alpha_{n,1} \theta^{t}_{local,n} +\alpha_{n,2} \theta^{t}_{C(n)}+ \alpha_{n,3} \theta^{t}_{global}.
     \end{split}
\end{equation}

After obtaining the embeddings, an additional linear layer or a dot multiplication of the user and item embeddings could be used to get the rating prediction.

\subsection{Server-Side Clustering Based Federation}
Our proposed server-side federation module performs three main functions: user clustering, user selection, and parameter aggregation. At each iteration, encrypted and privacy protected user/item embeddings and models are uploaded to the server by the users. 

$\bullet$~{\bf User Clustering.}
Based on user embeddings $E_{u,n}$, the server clusters users into $K$ groups. 
User $n$ belongs to cluster $C(n)$. We can use any commonly used clustering method such as K-means \cite{macqueen1967some}. Since the node representation $E_{u,n}$ is jointly learned from the attribute and collaborative information at each user, the representation is therefore enhanced.

$\bullet$~{\bf User Selection.}
To reduce the communication cost in critical conditions, our framework has an optional cluster-based user selection ability. Within each cluster, we can adaptively select a few random users, proportional to the cluster size, to participate in the model aggregation in each iteration.

$\bullet$~{\bf Parameter Aggregation.}
Our framework perform both network model aggregation and embedding aggregation. The user embedding is stored at the local user device but may get exchanged via the server without revealing the user identity. Item embeddings are shared and updated by all clients. For network models, the server will aggregate a global model $\theta^{t}_{global}$ (via a weighted sum of all participating users) and cluster-wise models $\theta^{t}_{C(n)}$ for cluster $C(n)$ (via a weighted sum of all participating users in the cluster). The global model $\theta^{t}_{global}$ and the cluster-level model $\theta^{t}_{C(n)}$ will be given to user $n$ for personalized recommendation.

\section{Experiments}
In this section, we show the effectiveness of our proposed framework over real-world datasets and compare our PerFedRec with existing baseline methods.

\subsection{Experiment Setup}
\textbf{Datasets.}
To evaluate our proposed framework, we conduct experiments on the three real-world datasets, whose statistics can be found in Table~\ref{dataset}. 
MovieLens\footnote{https://grouplens.org/datasets/movielens/} is a movie rating dataset that is widely used for evaluating recommendation algorithms. We use MovieLens-100k, which includes 100,000 user ratings. 
Yelp\footnote{https://www.yelp.com/dataset/challenge} is another widely-used benchmark dataset. 
Amazon-Kindle\footnote{https://jmcauley.ucsd.edu/data/amazon/} is from the Amazon review data to recommend e-books to users.


\begin{table}[h]
\centering
\caption{Dataset description}
\label{dataset}
\scalebox{0.95}[0.95]
{
\begin{tabular}{lcccc}
\hline
Dataset   & \# of user & \# of item & \# of rating & sparsity \\ \hline

MovieLens-100K & 943      & 1,682      & 100,000      & 93.70\%  \\ 
Yelp & 5,224      & 7,741      & 123,024      & 99.70\%  \\ 
Amazon-Kindle & 7,650      & 9,173      & 137,124      & 99.80\%  \\ \hline
\end{tabular}
}
\end{table}

\noindent\textbf{Baseline Methods.}
We compare our proposed PerFedRec solution with FedGNN \cite{wu2021fedgnn}, a federated recommendation framework using the popular FedAvg \cite{mcmahan2017communication} algorithm. 
Recall that PerFedRec is designed to improve FedAvg by clustering users and making personalized recommendations.
We also compare our model with a centralized version.
Since we conduct the experiments in a federated setting, we leave out some classical recommendation algorithms because it is hard to adapt them to the federated setting.

\noindent\textbf{Settings. }
In our experiments, we use a lightweight model named LightGCN \cite{he2020lightgcn} as the GNN model, and use the dot product to implement the rating predictor. The user and item embeddings and their hidden representations learned by graph neural networks are 64-dimensional. 
Following previous works \cite{deshpande2004item}, we apply the leave-one-out strategy for evaluation, and employ HR@K and NDCG@K to evaluate the performance. For each user, we use the last behavior for testing, the second to last for validation and the others for training. Similar to \cite{muhammad2020fedfast,elkahky2015multi}, we randomly sample 100 items that have no interaction with the user, and rank the test item among these items.
The number of users used in each round of model training is 128, and the default learning rate is 0.01. The hyper-parameters $\alpha_{n,1}$, $\alpha_{n,2}$, $\alpha_{n,3}$ are set to $\frac{1}{3}$.
The hyper-parameters (e.g., learning rate, dropout rate) of baseline methods are selected according to the best performance
on the validation set. 
The performance is averaged over 5 runs on the testing set.




\begin{table}[t]
\centering
\caption{Performance comparison (improv. represents the performance improvement of PerFedRec over FedAvg)}
\label{performance}
\scalebox{0.92}[0.92]{
\begin{tabular}{llllll}
\hline
                           \multicolumn{2}{l}{Model}         & Central & FedAvg & PerFedRec & Improv. \\ \hline
\multirow{4}{*}{MovieLens} & HR@10   & 0.6532  & 0.4475 & 0.6119    & 36.74\%     \\ 
                           & NDCG@10 & 0.4337  & 0.2462 & 0.4409    & 79.08\%     \\ \cline{2-6} 
                           & HR@20   & 0.8165  & 0.6405 & 0.7402    & 15.57\%     \\ 
                           & NDCG@20 & 0.4740  & 0.2954 & 0.4707    & 59.34\%     \\ \hline
\multirow{4}{*}{Yelp}      & HR@10   & 0.6497  & 0.4422 & 0.6058    & 36.99\%     \\ 
                           & NDCG@10 & 0.4099  & 0.2487 & 0.4040    & 62.44\%    \\ \cline{2-6} 
                           & HR@20   & 0.8168  & 0.6437 & 0.7151    & 11.09\%     \\ 
                           & NDCG@20 & 0.4531  & 0.3000 & 0.4317    & 43.90\%     \\ \hline
\multirow{4}{*}{Kindle}    & HR@10   & 0.6382  & 0.3936 & 0.4514    & 14.68\%     \\ 
                           & NDCG@10 & 0.4102  & 0.2366 & 0.3111    & 31.49\%     \\ \cline{2-6} 
                           & HR@20   & 0.7742  & 0.5218 & 0.5548    & 6.32\%      \\  
                           & NDCG@20 & 0.4448  & 0.2701 & 0.3379    & 25.10\%     \\ \hline
\end{tabular}}
\end{table}

\begin{table}[]
\centering
\caption{Performance under different \# of clusters $K$ on MovieLens dataset}
\label{cluster}
\scalebox{0.95}[0.95]{
\begin{tabular}{lllll}
\hline
   \# of clusters    & \multicolumn{1}{l}{HR@10}  & \multicolumn{1}{l}{NDCG@10} & \multicolumn{1}{l}{HR@20}  & NDCG@20 \\ \hline
$K=5$  & \multicolumn{1}{l}{0.6119} & \multicolumn{1}{l}{0.4409}  & \multicolumn{1}{l}{0.7402} & 0.4707  \\ 
$K=10$ & \multicolumn{1}{l}{0.6214} & \multicolumn{1}{l}{0.4492}  & \multicolumn{1}{l}{0.7470} & 0.4799  \\ 
$K=20$ & \multicolumn{1}{l}{0.6087} & \multicolumn{1}{l}{0.4404}  & \multicolumn{1}{l}{0.7328} & 0.4699  \\ \hline
\end{tabular}
}
\end{table}

\subsection{Performance Evaluation}
Table~\ref{performance} shows the performance of all three methods on three datasets. 
The centralized method achieves the best results in almost all scenarios, and 
FedAvg achieves the worst results on all scenarios since it ignores the feature information and does not provide personalized recommendation.

The performance of our proposed PerFedRec is close to the centralized method in most cases.
Compared to FedAvg, our proposed PerFedRec achieves an improvement of 29.47\% in terms of HR@10 on average over all three datasets and 57.67\% in terms of NDCG@10 on average over all three datasets.
The improvement (43.79\% on average) is the most significant on the MovieLens dataset.
The improvement is smaller for sparser datasets such as Kindle while still achieving 19.40\% on average. 
In particular, the improvement on the Kindle dataset, which does not have external feature information, shows the importance of personalized recommendations.




\subsection{Model Analysis}
We discuss the impacts of a key hyperparameter: the number of clusters $K$ during model training. Table~\ref{cluster} indicates that the performance is relatively stable under varying hyperparameters, which reduces the burden of hyperparameter tuning. 

\subsection{Ablation Study}
We conduct ablation study to evaluate how much each module of our proposed framework contributes to the performance. Specifically, PerFedRec-Variation 1 uses no personalized recommendation,
PerFedRec-Variation 2 uses no feature information,
and PerFedRec-Variation 3 uses no user clustering.
Table~\ref{ablation} shows that the largest improvement comes from personalized recommendation. Moreover, incorporating feature information brings an obvious improvement on the performance. Finally, compared to Variation 3 (no user clustering), PerFedRec has negligible performance degradation while reducing communication cost by user clustering. 

\begin{table}[]
\centering
\caption{Ablation study on MovieLens dataset}
\label{ablation}
\scalebox{0.85}[0.85]{
\begin{tabular}{lllll}
\hline
    Model      & \multicolumn{1}{l}{HR@10}  & \multicolumn{1}{l}{NDCG@10} & \multicolumn{1}{l}{HR@20}  & NDCG@20 \\ \hline
PerFedRec-Variation 1 & \multicolumn{1}{l}{0.4464} & \multicolumn{1}{l}{0.2497}  & \multicolumn{1}{l}{0.6405} & 0.2965  \\
PerFedRec-Variation 2 & \multicolumn{1}{l}{0.6087} & \multicolumn{1}{l}{0.4340}  & \multicolumn{1}{l}{0.7381} & 0.4677  \\ 
PerFedRec-Variation 3 & \multicolumn{1}{l}{0.6102} & \multicolumn{1}{l}{0.4453} & \multicolumn{1}{l}{0.7359} & 0.4766 \\ 
PerFedRec      & \multicolumn{1}{l}{0.6119} & \multicolumn{1}{l}{0.4409}  & \multicolumn{1}{l}{0.7402} & 0.4707  \\ \hline
\end{tabular}
}
\end{table}

\section{Conclusion}
In this paper, we highlight the importance of user clustering in personalized federated recommendations and propose a novel personalized federated recommendation framework. The proposed framework jointly learns user representations from collaborative and attribute information via GNNs, clusters similar users, and obtains personalized recommendation models by combining the user-level, cluster-level, and global models.
To alleviate the communication burden, we propose a sampling strategy to select representative clients from each cluster for model federation.
Experiments on three real-world datasets demonstrate that our proposed framework achieves superior performance for federated recommendation.

\begin{acks}
This work was supported in part by the Changsha Science and Technology Program International and Regional Science and Technology Cooperation Project under Grants kh2201026, the Hong Kong RGC grant ECS 21212419, the Technological Breakthrough Project of Science, Technology and Innovation Commission of Shenzhen Municipality under Grants JSGG20201102162000001, InnoHK initiative, the Government of the HKSAR, Laboratory for AI-Powered Financial Technologies, the Hong Kong UGC Special Virtual Teaching and Learning (VTL) Grant 6430300, and the Tencent AI Lab Rhino-Bird Gift Fund.
\end{acks}

\bibliographystyle{ACM-Reference-Format}
\balance
\bibliography{sample-base}

\end{document}